# Electron wavefunction probing in room-temperature semiconductors: direct observation of Rabi oscillations and self-induced transparency

Amir Capua[1], Ouri Karni[1], Gadi Eisenstein[1], Johann Peter Reithmaier[2]

[1] Electrical Engineering Department, Technion, 32000 Haifa, Israel

[2] Institute of Nanostructure Technologies and Analytics, Technische Physik, CINSaT, University of Kassel, 34132 Kassel, Germany

**Quantum coherent light-matter interactions have been at the forefront of scientific interest since the fundamental predictions of Einstein and the later work of Rabi[1-3]. Direct observation of quantum coherent interactions entails probing the electronic wavefunction which requires that the electronic state of the matter does not de-phase during the measurement, a condition that can be satisfied by lengthening the coherence time or by shortening the observation time. The short de-phasing time in semiconductors has dictated that all coherent effects reported to date have been recorded directly only at cryogenic temperatures[4-17]. Here we report on the first direct electronic wavefunction probing in a room-temperature semiconductor. Employing an ultrafast characterization scheme we have demonstrated Rabi oscillations and self-induced transparency in an electrically driven, room-temperature semiconductor laser amplifier, revealing the most intimate details of the light-matter interactions seen to date. The ability to employ quantum effects in solid-state media, which operate at elevated temperatures, will finally bring true quantum mechanical concepts into the realm of practical devices.**

The induction and probing of quantum coherent phenomena are crucial for the understanding of light-matter interactions. The essence of such coherent interactions is that the electronic wavefunction and the electromagnetic field evolve together. This requires that the electronic states of the matter persist longer than the duration of the interaction. Several coherent phenomena were demonstrated at elevated temperature atomic vapors[18-21] where the weak interactions within dilute atom ensembles ensure long coherence times, up to a few milli seconds. In solids, particularly in optically active semiconductors, the atomic densities are large and scattering processes shorten the coherence lifetimes significantly. Experiments seeking observation of quantum coherent phenomena in semiconductors have therefore always been performed at cryogenic temperatures[4-17]. At room-temperature, the coherent lifetime in semiconductors is very short, of the order of 0.5-1 ps[22] and hence direct observations of coherent phenomena require ultrafast characterization schemes.

In this paper, we report on the first room-temperature direct observation of quantum coherent phenomena in a semiconductor medium. We use a cross frequency resolved optical gating (X-FROG) technique[23] with a temporal resolution of a few femto-seconds to measure the time resolved amplitude and phase of a 200 fs pulse after propagation in an InAs/InP quantum dash (wire-like)[24] laser amplifier. Operating in the gain regime, we demonstrate Rabi oscillations while under absorbing conditions, we observe clearly self-induced transparency. The experimental observations are consistent with a detailed Maxwell-Schrödinger model of a two-level system fed by a carrier reservoir.

In order to understand the principles governing quantum coherent interactions induced by short pulses which propagate in an electrically pumped laser amplifier, we analyze first the case of a moderately intense pulse, which causes the gain to saturate into transparency, as described in Fig. 1.

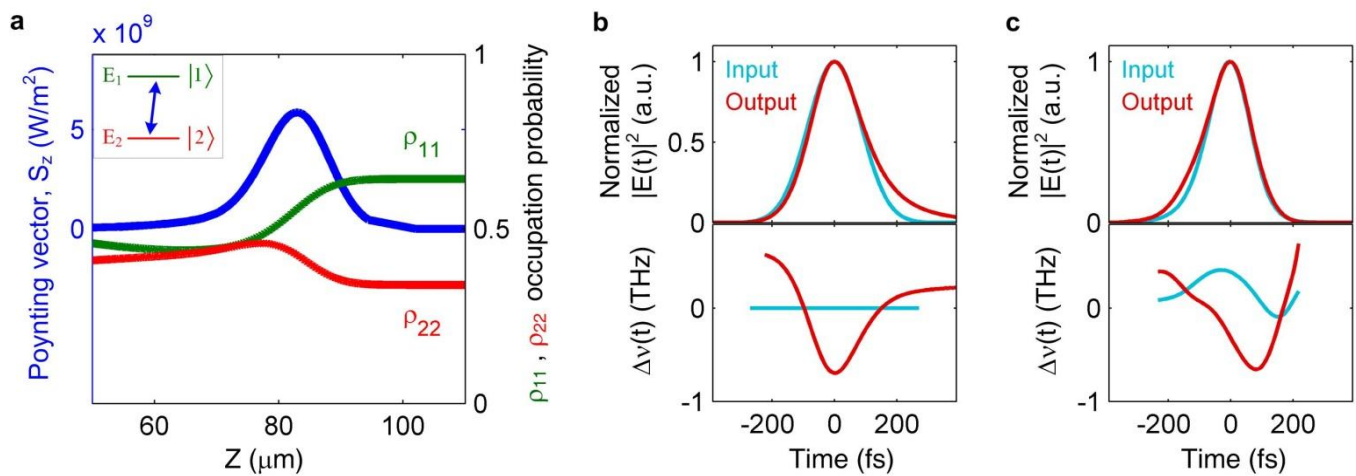

Figure 1

**Figure 1| Classical gain saturation regime.** Propagation of a ~ 200 fs pulse whose area is π/2. The cycle of the Rabi oscillation is much longer than the pulse width. (a) Simulated spatial distribution of $\rho_{11}$ and $\rho_{22}$ along a 60 μm section of the laser waveguide under conditions of 10 dB gain together with the π/2 pulse which is placed 80 μm from the input facet. Shown in blue is the envelope of the pulse Poynting vector along the direction of propagation, $S_z$. The distributions of $\rho_{11}$ and $\rho_{22}$ exhibit a transition from gain to transparency due to the pulse, and gain recovery long after the pulse has passed. The inset shows schematically the ground state represented by the energy levels $E_1$ and $E_2$. (b) Simulated normalized pulse intensity, $\left|E(t)\right|^2$, and instantaneous frequency shift, $\Delta\nu(t)$, at the input and output of the laser amplifier. (c) Measured normalized pulse intensity and instantaneous frequency shift at the input and output of laser amplifier. Both the experiment and measurement show no significant pulse distortion. The time dependent instantaneous frequency exhibits typical behavior during amplification: red shift during the leading edge of the pulse followed by a frequency increase during the trailing edge. This functional form is termed "valley" to identify gain events. Note that the input pulse in the experiment is initially chirped.

Fig. 1a shows results obtained from a Maxwell-Schrödinger model of the spatially dependent occupation probability amplitudes in the upper and lower levels of the ground state (marked in the inset as the energy levels $E_1$ and $E_2$) $\rho_{11}$ and $\rho_{22}$, respectively, along a 60 $\mu$m section of the laser waveguide. Also shown is the envelope of the pulse Poynting vector in the direction of propagation. Immediately behind the pulse, the medium is transparent ($\rho_{11} = \rho_{22}$) while it exhibits population inversion ($\rho_{11} > \rho_{22}$) wherever the pulse has not arrived. The transition takes place during the pulse while far behind the pulse, the probabilities show initial recovery. The saturation effect (Fig. 1a) can alternatively be described as a coherent phenomenon where the transition from gain to transparency is induced by a pulse whose area[2] is π/2. The Rabi frequency, $\Omega = \mu E / \hbar$ of this moderate intensity π/2 pulse yields a low $\Omega$ for which the Rabi-oscillation period is longer than the pulse duration and hence $\rho_{11}$ and $\rho_{22}$ develop monotonically during the pulse and exhibit no oscillatory features.

The calculated output pulse intensity profile and instantaneous frequency shift relative to its carrier frequency are shown in Fig. 1b together with those of the input pulse. Since saturation is moderate, the pulse profile experiences no significant distortion. The behavior of the instantaneous frequency is determined by the fact that changes in the refractive index are inversely proportional to carrier density changes. In the gain regime, carrier depletion causes a red shift during the pulse leading edge followed by a recovery during the trailing end. Hence, the instantaneous frequency exhibits a "valley" denoting the occurrence of a gain event. Experimental characterization, using the X-FROG system, of a π/2 pulse after traversing the laser amplifier is presented in Fig. 1c. The intensity profile and the time dependent instantaneous frequency confirm the predictions of Fig. 1b.

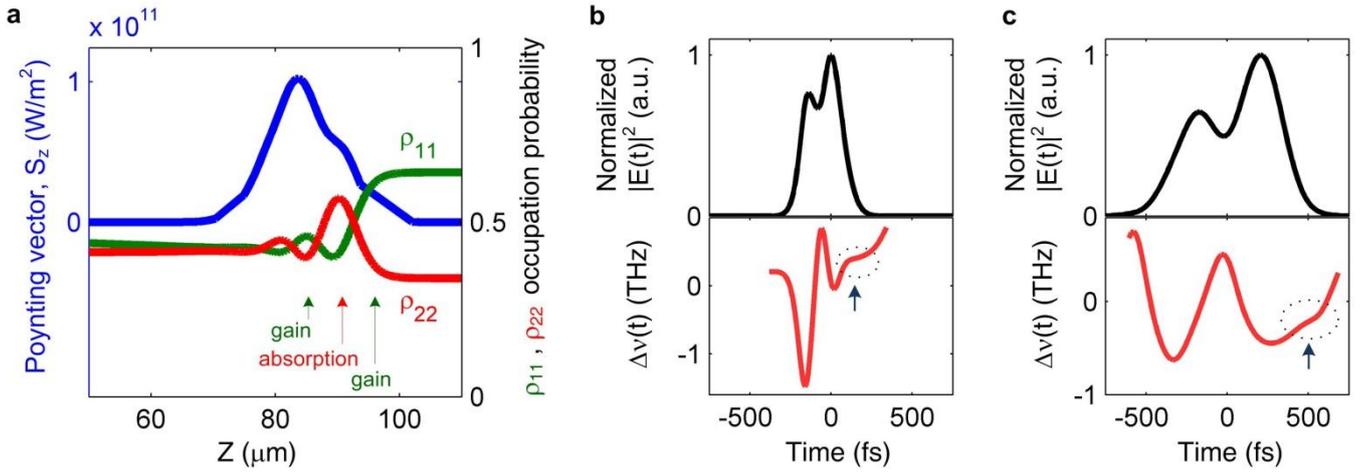

Figure 2

**Figure 2| Rabi oscillation dominated propagation. Propagation of an intense ~ 200 fs pulse whose area is 4.4π. The cycle of the Rabi oscillation is smaller than the pulse duration. (a) Simulated spatial distribution of $\rho_{11}$ and $\rho_{22}$ along a 60 μm section of the laser waveguide under conditions of 10 dB gain together with a 4.4π pulse which is placed 80 μm from the input facet. The distributions of $\rho_{11}$ and $\rho_{22}$ exhibit clear Rabi oscillations during the pulse. Population inversion exists ahead of the pulse while far behind the pulse, the medium is transparent and eventually the gain recovers. (b) Simulated output pulse. The normalized intensity profile comprises two peaks while the time dependent instantaneous frequency shows two valleys indicating two distinct gain events. (c) Measured output pulse. The two-peak normalized intensity profile and the two valleys representing two gain events are clearly seen, consistent with the simulation results in (b). Arrows are added to guide the reader's eye to the fine details in the calculated instantaneous frequency trace which are revealed also in the experimental result. These fine details signify the electronic wavefunction and are reproduced throughout the paper.**

Increasing the pulse area and correspondingly the Rabi frequency leads to Rabi oscillation periods, which are shorter than the duration of the pulse. This causes a complete inversion of the occupation probabilities known as Rabi flopping. A simulation of this process for pulses with an area of 4.4π is shown in Fig 2a. Here, the conditions sensed by different parts of the pulse alternate between gain ($\rho_{11} >$ $\rho_{22}$) and coherent absorption ($\rho_{11} < \rho_{22}$). At locations far behind and ahead of the pulse, the conditions are the same as in the "classical gain saturation" case (Fig. 1a), transparency and population inversion, respectively. The resultant intensity profile at the output exhibits a double peaked pulse as seen in Fig. 2b. The corresponding time dependent instantaneous frequency, shown also in Fig. 2b, comprises two valleys, which signify two distinct amplification events occurring within the duration of the pulse. The calculated characteristics of this intense, 4.4π, pulse were confirmed by X-FROG measurements as described in Fig. 2c. The difference in the output pulse width between Fig. 2b and Fig. 2c stems from two-photon absorption and the gain broadening inhomogeneity, both of which are not accounted for in the model. Nevertheless, the main oscillatory behavior in both the intensity profiles and the time dependent instantaneous frequency traces are clearly seen and testify to the occurrence of distinct Rabi oscillations.

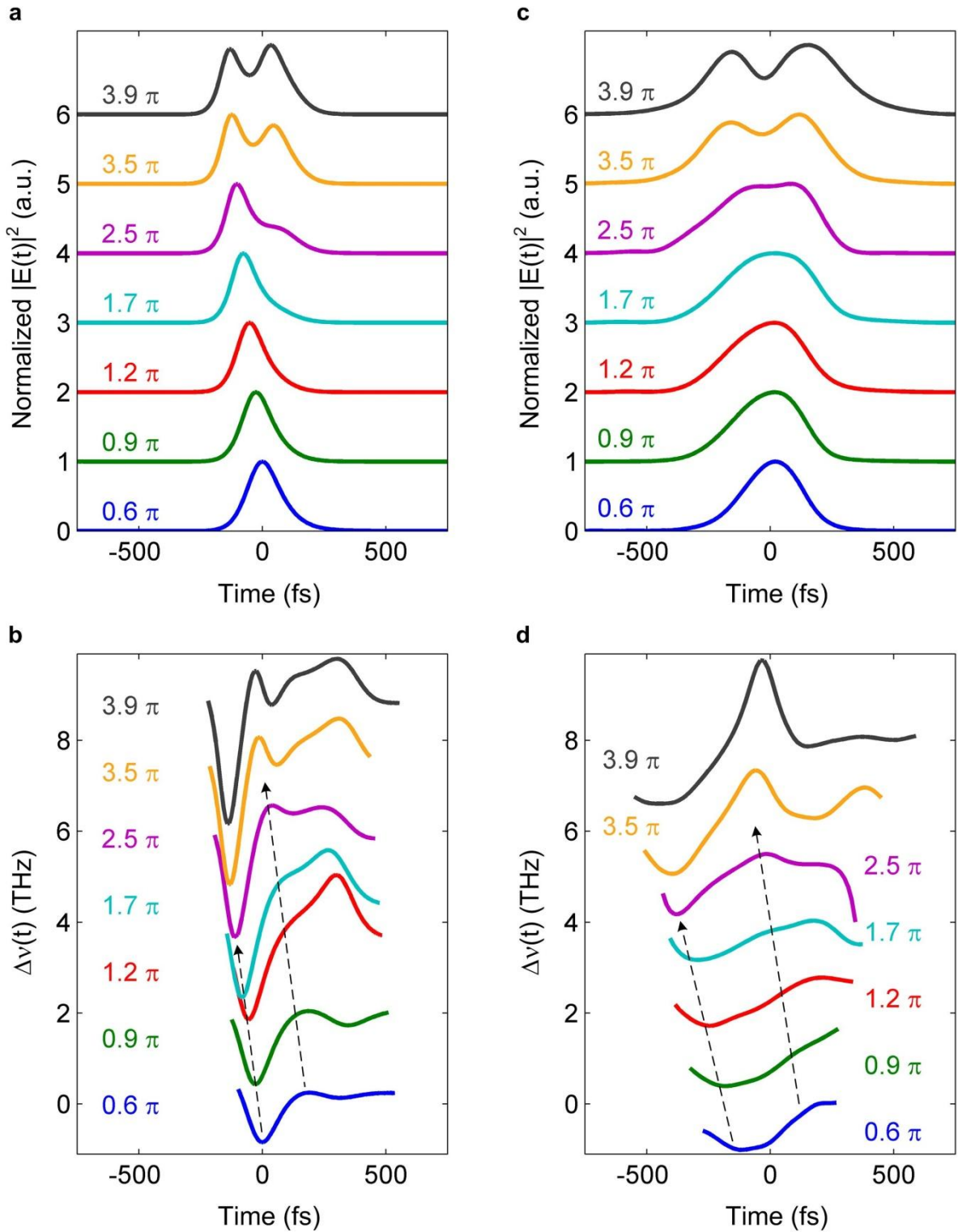

Figure 3

**Figure 3| Area dependence of the output pulses.** The input pulse energy is modified to vary the pulse area from 0.6π to 3.9π while the gain is 10 dB. (a) and (b) Simulated pulse-area dependent normalized output intensity profiles and time dependent instantaneous frequencies. Rabi oscillations are observed for input pulse areas above 2.5π. Pulse break-up and the corresponding second valley emerge gradually. The first gain saturation event

**shifts to earlier times as the pulse area increases. (c) and (d) Measured pulse-area dependent normalized output intensity profiles and time dependent instantaneous frequencies. The measurements are displayed on an absolute common time axis. The measurements show fine details all of which are consistent with the predicted behavior in (a) and (b). Arrows are added to guide the reader's eye to the evolution. Traces are shifted for clarity.**

The transition from classical saturation to a regime where Rabi oscillations dominate was further investigated by gradually increasing the pulse area, either by using larger input pulse energies or by raising the bias. The dependence on input pulse area is described in Fig. 3 which shows simulated (3a and 3b) and measured (3c and 3d) results. The singly peaked low power pulse profile evolves gradually into the doubly peaked shape observed for input pulse areas larger than 2.5π. The time dependent instantaneous frequency evolves correspondingly from the conventional single valley shape at low pulse energies to a double valley functional form. The first valley, which denotes classical gain saturation, shifts to earlier times as the pulse area is increased since saturation occurs obviously earlier for more intense pulses. The simulations and the experiments are consistent with each other where even the finest measured features in the instantaneous frequency traces are accurately predicted by the model.

An alternative way to increase the pulse area is to operate with larger gain levels while keeping the input energy constant. Figure 4 shows bias dependent responses for an input pulse whose area is 3.2π. A double-peaked intensity and an instantaneous frequency profile with two valleys are seen indicating that Rabi oscillations take place. As the bias increases, these two signatures of the coherent interaction become more pronounced. In particular, the oscillation lasts for a longer part of the pulse duration as evident by the second cycle in the instantaneous frequency traces. Once more, the simulation predicts all the details measured by the X-FROG system.

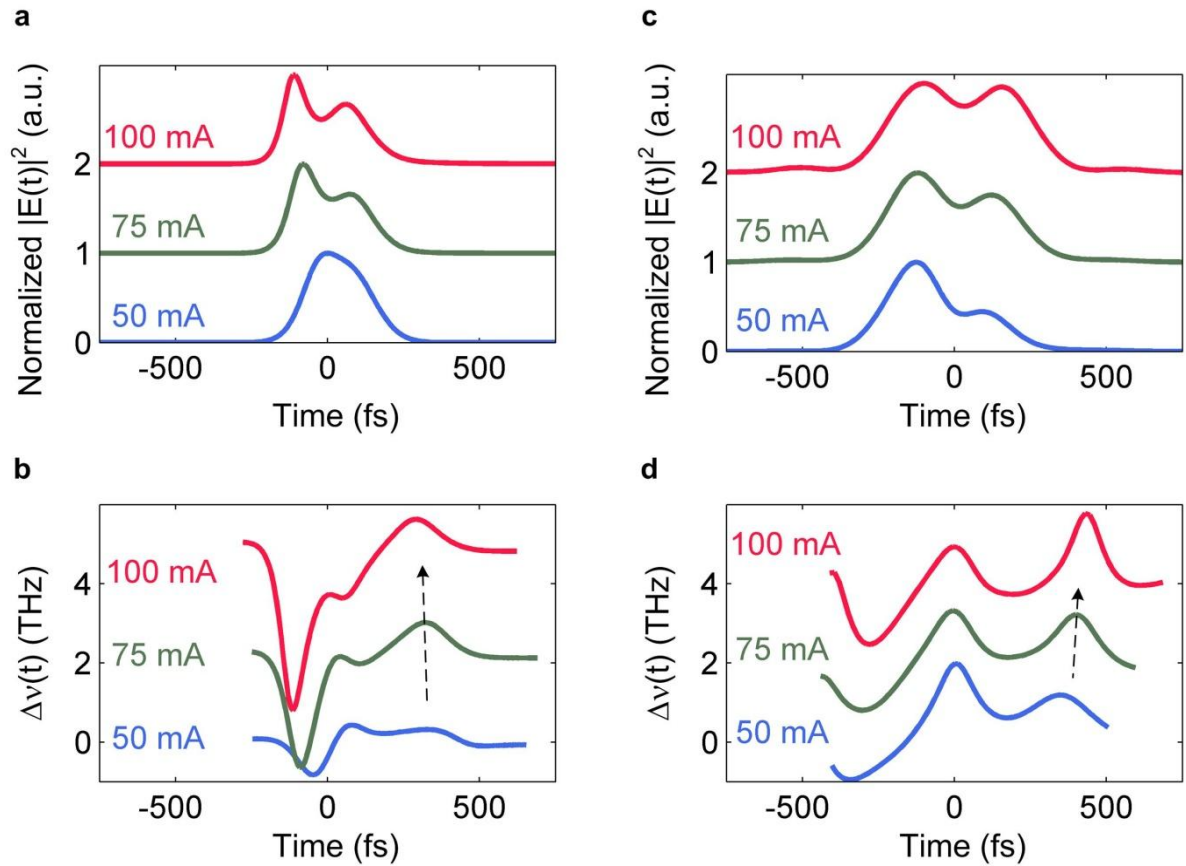

Figure 4

**Figure 4| Bias dependence of the output pulses. Output pulses for bias levels of 50 mA, 75 mA and 100 mA corresponding to gain levels 5 dB, 7.5 dB and 10 dB, respectively. The input pulse area was 3.2π. Simulated results (a) and (b) show the predicted evolutions, which are confirmed by the measurements (c) and (d). Arrows are added to guide the reader's eye to the evolution. Traces are shifted for clarity.**

The complimentary effect to the Rabi oscillations is self-induced transparency[6,8,14, 18,21,25]. In this case, the effective two-level system is prepared, prior to the arrival of the electromagnetic field, in its lower state. This is achieved here by applying zero bias to the device so it operates in the absorption regime. Self-induced transparency means that an intense pulse may co-evolve with the medium pumping it beyond the transparency point into the gain regime. This requires once more that the period of the Rabi cycle is shorter than the pulse width. As illustrated in the calculated spatial distribution of $\rho_{11}$ and $\rho_{22}$ (Fig. 5a), the leading edge of the pulse is absorbed, the following central part undergoes at the same time amplification and the trailing edge is absorbed. Pulse propagation of this kind is well known to result in pulse compression. Actually, Fig. 5a shows that an additional gain event takes place in the trailing part of the pulse.

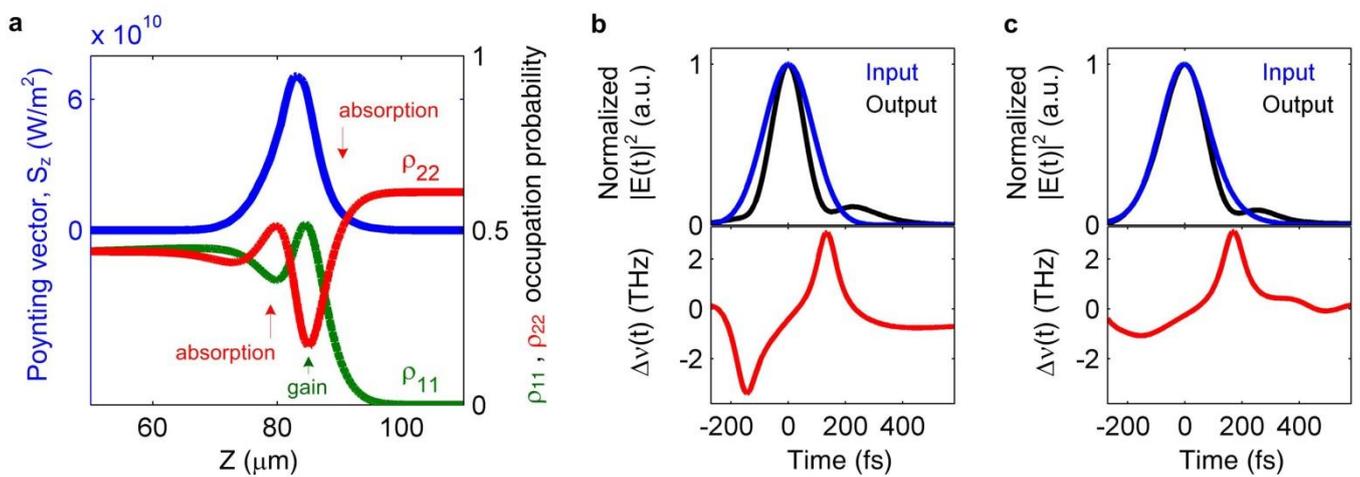

Figure 5

**Figure 5| Self-induced transparency.** Simulations and measurements of a pulse whose area is 3.6π propagating under zero bias conditions. The cycle of the Rabi oscillation is shorter than the pulse duration. (a) Simulated spatial distribution of $\rho_{11}$ and $\rho_{22}$ along a 60 μm section of the laser waveguide under conditions of zero bias together with a 3.6π pulse which is placed 80 μm from the input facet. During the pulse, the medium flips from absorption to gain and back to absorption. The unperturbed amplitude probability $\rho_{22}$ is smaller than unity due to occupation of reservoir states according to the principle of detailed balance (see methods). The interaction optically pumps the medium so that behind the pulse it approaches transparency. (b) Simulated normalized output intensity and the time dependent instantaneous frequency. The intensity profile shows symmetrical pulse compression. (c) Measured normalized output intensity and time dependent instantaneous frequency showing asymmetrical pulse compression (on the trailing edge only) as well as traces of the following second oscillation. The time dependent instantaneous frequency shows features which agree with the simulated results.

In the zero bias case, the free carrier density is low and its contribution to the refractive index changes is negligible compared to the one in the gain regime.

Simulating this case requires therefore to modify the refractive index dependence on carrier density. This dependence was thoroughly studied, theoretically and experimentally, by Zilkie[26]. At zero bias, two-photon absorption and stimulated transition heating govern the refractive index dynamics. The combined effects manifest themselves in an opposite index dependence on carrier density, compared to the gain regime. Using this dependence, we calculate the output pulse properties. The intensity profile of Fig. 5b shows the expected pulse compression together with the time resolved instantaneous frequency. The corresponding X-FROG measurements are presented in Fig. 5c. While the simulation shows symmetric pulse compression, the experiment shows a clear compression on the trailing edge and a very slight compression on the leading edge. Additionally, a second trailing excitation in the intensity profile is observed which also appears in the simulations and indicates that a second oscillation cycle was initiated. The measured time resolved instantaneous frequency, Fig. 5c, is consistent with the simulated result exhibiting a red shift during the pulse peak, which is followed by a sharp frequency increase during the trailing edge.

To summarize, we have used X-FROG characterization to demonstrate for the first time direct electron wavefunction probing in a semiconductor. Furthermore, the experiments we report were performed at room temperature. The high resolution phase-sensitive measurement technique we have employed revealed the finest details of the coherent interaction between light and the electronic wavefunction enabling to demonstrate the two fundamental processes of Rabi oscillations and self-induced transparency using 200 fs wide pulses which propagated in an electrically driven InAs/InP quantum dash laser amplifier. A comprehensive simulation in which the principle interaction was governed only by the Schrödinger and two Maxwell equations, is consistent with all the measured results proving that the medium behaves effectively as a two-level system. Our work sheds light on the nature of quantum coherent interactions between electromagnetic waves and the collective wavefunction of a large ensemble of atoms[27] and stimulates future research into many fundamental principles such as the Bloch theorem and the area theorem of McCall and Hann[25]. Most importantly, our work opens a window for the exploitation of fascinating quantum phenomena in room-temperature semiconductor devices that enact electronic wavefunction manipulation. Such devices will lead to practical uses of quantum effects such as quantum storage, quantum key distribution, entanglement purification, quantum lithography etc. and hence is expected to have a major impact on communication, computation, imaging and sensing.

## Methods

**QDash Laser amplifier** The laser utilized in the experiments was 1.5 mm long. Its gain section comprised six InAs/InP QDash layers grown by MBE[28]. The end facet reflectivities were reduced to ~ 0.04% by applying a broadband multi-layer anti-reflection coating. Estimation of the pulse area assumed a wave-guide cross-section of 9 $\mu m^2$ and a material index of 3.46.

**Maxwell-Schrödinger model** The co-evolution of the electronic wavefunction and the electromagnetic wave were calculated by applying the Maxwell and Schrödinger equations on a two-level system which was described under the density matrix framework using the dipole approximation. The two-level system was coupled to a carrier reservoir by capture and escape processes to account for the so-called nanostructure in-a-well arrangement. The carrier dynamics were described by a set of common rate equations whose driving term is an electrical bias. The model calculated the hole and electron densities as well as carrier diffusion along the propagation axis. All the equations were solved self-consistently using the finite difference time domain (FDTD) technique. The calculations do not invoke the usual rotating wave approximation but rather preserve the oscillatory nature of the electronic wavefunction and the electromagnetic wave. The fundamental equations that describe the light and matter interaction are given by [29,30]:

$$\begin{cases} \dfrac{d\rho_{11}}{dt} = \Lambda_e - \gamma_{11}\rho_{11} + \dfrac{\mu}{i\hbar}\left(\rho_{12} - \rho_{21}\right)E(t) \\[2mm] \dfrac{d\rho_{22}}{dt} = -\Lambda_h - \gamma_{22}\rho_{22} - \dfrac{\mu}{i\hbar}\left(\rho_{12} - \rho_{21}\right)E(t) \\[2mm] \dfrac{d\rho_{12}}{dt} = -\left(i\omega + \gamma_{12}\right)\rho_{12} - \dfrac{i\mu}{\hbar}\left(\rho_{11} - \rho_{22}\right)E(t) \end{cases}$$

With $\rho_{11}$, $\rho_{22}$ being the occupation probability of the upper and lower energy level of the two-level system, respectively and where $\rho_{12}$, $\rho_{21}$ are the carrier coherence terms. $\mu$, $\omega$, $\gamma_{11}$, $\gamma_{22}$ are the dipole moment, angular frequency of the transition and lifetimes of the upper and lower levels respectively while $\gamma_{12}$ is the de-phasing rate. The coupling to the electron and hole carrier reservoirs is expressed through the rates $\Lambda_e$ and $\Lambda_h$.

The phenomenological dependence of the index on the carrier concentration was expressed by:

$$n^2 = \varepsilon_{r_0} - \varepsilon\left(N_{res}, \rho_{11}\right) = \varepsilon_{r_0} - C_{res}N_{res} - C_{2lev}\rho_{11}$$

with $\varepsilon_{r_0}$ being the background index and $N_{res}$ the reservoir carrier density. $C_{res}$ and $C_{2lev}$ describe the index dependence on the reservoir and two-level populations.

The simulation assumes that the reflection of the end-facets was zero. 200 fs wide (full-width at half-maximum) transform limited Gaussian pulses were applied at the input.

**X-FROG setup** The X-FROG system used pulses generated by a tunable optical parametric oscillator (Spectra-Physics OPAL) emitting 200 fs pulse at 82 MHz with a maximum average power of 250 mW. The spectral information was obtained using a handheld spectrometer (Ocean Optics USB 4000) and a spectral marginal was applied based on an optical spectrum analyzer (Ando AQ-6317). Optical gating was performed using a 0.5 mm long LiNbO$_3$ crystal. Coupling in and out of the device employed two NA 0.63 objective lenses whose losses were considered in the evaluation of the pulse area. Data retrieval was performed with commercial software package (Femtosoft Technologies). Typical convergence errors were below 0.002. Simple modifications transform the X-FROG system into a standard FROG set up which is used for pulse pre-characterization. The X-FROG scheme is advantageous over the simple FROG technique in two ways; it offers an improved sensitivity (by up to three orders of magnitude in our case) and the extracted data can be placed on an absolute common time axis.

## References

1. Rabi, I. I. Space quantization in a gyrating magnetic field. *Phys. Rev.* **51,** 652–654 (1937).

2. Hopf, F. A. & Scully, M. O. Theory of an inhomogeneously broadened laser amplifier. *Phys. Rev.* **179,** 399–416 (1969).

3. Lamb, G. L. JR. Analytical descriptions of ultrashort optical pulse propagation in a resonant medium. *Rev. Mod. Phys.* **43,** 99-124 (1971).

4. Cundiff, S.T. et al. Rabi flopping in semiconductors. *Phys. Rev. Lett.* **73,** 1178–1181 (1994).

5. Fürst, C., Leitenstorfer, A., Nutsch, A., Tränkle, G. & Zrenner, A. Ultrafast Rabi oscillations of free-carrier transitions in InP. *Phys. Stat. Sol. B* **204**, 20–22 (1997).

6. Giessen, H. et al. Self-induced transmission on a free exciton resonance in a semiconductor. *Phys. Rev. Lett.* **81,** 4260–4263 (1998).

7. Schülzgen, A. et al. Direct observation of excitonic Rabi oscillations in semiconductors. *Phys. Rev. Lett.* **82,** 2346–2349 (1999).

8. Jütte, M. & von der Osten, W. Self-induced transparency at bound excitons in CdS . *J. Luminescence* **83-84**, 77-82 (1999).

9. Cole, B. E., Williams, J. B., King, B. T., Sherwin, M. S. & Stanley, C. R. Coherent manipulation of semiconductor quantum bits with terahertz radiation. *Nature* **410,** 60-63 (2001).

10. Stievater, T. H. et al. Rabi oscillations of excitons in single quantum dots. *Phys. Rev. Lett.* **87,** 133603 (2001).

11. Htoon, H. et al. Interplay of Rabi oscillations and quantum interference in semiconductor quantum dots. *Phys. Rev. Lett.* **88,** 087401 (2002).

12. Zrenner, A. et al. Coherent properties of a two-level system based on a quantum-dot photodiode. *Nature* **418,** 612-614 (2002).

13. Borri, P. et al. Rabi oscillations in the excitonic ground-state transition of InGaAs quantum dots. *Phys. Rev. B* **66**, 081306(R) (2002).

14. Schneider, S. et al. Self-induced transparency in InGaAs quantum-dot waveguides. *Appl. Phys. Lett.* **83,** 3668 (2003).

15. Flagg, E. B. et al. Resonantly driven coherent oscillations in a solid-state quantum emitter. *Nature Phys.* **5,** 203 - 207 (2009).

16. Choi, H. et al. Ultrafast Rabi flopping and coherent pulse propagation in a quantum cascade laser. *Nature Photon*. **4,** 706–710 (2010).

17. Zecherle, M. et al. Ultrafast few-fermion optoelectronics in a single self-assembled InGaAs/GaAs quantum dot. *Phys. Rev. B* **82,** 125314 (2010).

18. Gibbs, H. M. & Slusher, R. E. Sharp-line self-induced transparency. *Phys. Rev. A* **6,** 2326–2334 (1972).

19. Boller, K.-J., Imamoglu, A. S. & Harris, E. Observation of electromagnetically induced transparency *Phys. Rev. Lett.* **66,** 2593–2596 (1991).

20. Lezama, A., Barreiro, S. & Akulshin, A. M. Electromagnetically induced absorption *Phys. Rev. A* **59,** 4732–4735 (1999).

21. Won-Kyu, L. et al. Self-induced transparency in samarium atomic vapor under condition of high temperature and high density. *Jpn. J. Appl. Phys.* **41,** 5170–5176 (2002).

22. Bayer, M. & Forchel, A. Temperature dependence of the exciton homogeneous linewidth in $In_{0.60}Ga_{0.40}As$/GaAs self-assembled quantum dots. *Phys. Rev. B* **65,** 041308(R) (2002).

23. Trebino, R. *Frequency-Resolved Optical Gating: The Measurement Of Ultrashort Laser Pulses.* (Kluwer Academic Publishers, Norwell, 2002).

24. Dery, H. & Eisenstein, G. Self consistent rate equations of self assembly quantum wire lasers. *IEEE J. Quantum Electron.* **40,** 1398–1409 (2004).

25. McCall, S. L. & Hahn, E. L. Self-induced transparency. *Phys. Rev.* **183,** 457-485 (1969).

26. Zilkie, A.J. et al. Time-resolved linewidth enhancement factors in quantum dot and higher-dimensional semiconductor amplifiers operating at 1.55 μm. *J. Lightw. Technol.* **26,** 1498 - 1509 (2008).

27. De Riedmatten, H., Afzelius, M., Staudt, M. U., Simon, C., & Gisin N. A solid-state light–matter interface at the single-photon level. *Nature* **456,** 773-777 (2008).

28. Reithmaier, J.P. et al. InP based lasers and optical amplifiers with wire-/dot-like active regions, *J. Phys. D* **38,** 2088-2102 (2005).

29. Sargent III, M., Scully, M. O. & Lamb, W. E. Jr. Laser Physics (Westview Press, Boulder, 1978).

30. Yao, J., Agrawal, G. P., Gallion, P. & Bowden, C. Semiconductor laser dynamics beyond the rate-equation approximation. *Opt. Comm.* **119,** 246 -255 (1995).